\documentclass{elsart}
\usepackage[square,comma]{natbib}
\usepackage{graphicx}
\usepackage{pxfonts}
\usepackage{lineno}
\usepackage{amssymb}
\usepackage{footnote}%!!!
\usepackage{multirow}
\usepackage{varwidth}
\usepackage{array}
\usepackage{epstopdf}
\usepackage{url}
\usepackage{color} % for red text

\journal{}

\begin{document}
\thispagestyle{empty}
\begin{Large}
\textbf{DEUTSCHES ELEKTRONEN-SYNCHROTRON}

\textbf{\large{Ein Forschungszentrum der
Helmholtz-Gemeinschaft}\\}
\end{Large}

DESY 17-093

June 2017

\begin{eqnarray}
\nonumber &&\cr \nonumber && \cr \nonumber &&\cr
\end{eqnarray}
\begin{eqnarray}
\nonumber
\end{eqnarray}
\begin{center}
\begin{Large}
\textbf{On Radiation Emission from a Microbunched Beam with Wavefront Tilt and Its Experimental Observation}
\end{Large}
\begin{eqnarray}
\nonumber &&\cr \nonumber && \cr
\end{eqnarray}

\begin{large}
Gianluca Geloni,
\end{large}
\textsl{\\European XFEL GmbH, Hamburg}
\begin{large}

Vitali Kocharyan and Evgeni Saldin
\end{large}
\textsl{\\Deutsches Elektronen-Synchrotron DESY, Hamburg}
\begin{eqnarray}
\nonumber
\end{eqnarray}
\begin{eqnarray}
\nonumber
\end{eqnarray}
ISSN 0418-9833
\begin{eqnarray}
\nonumber
\end{eqnarray}
\begin{large}
\textbf{NOTKESTRASSE 85 - 22607 HAMBURG}
\end{large}
\end{center}
%\end{widetext}
\clearpage
\newpage

\begin{frontmatter}

% Title, authors and addresses

% use the thanksref command within \title, \author or \address for footnotes;
% use the corauthref command within \author for corresponding author footnotes;
% use the ead command for the email address,
% and the form \ead[url] for the home page:
% \title{Title\thanksref{label1}}
% \thanks[label1]{}
% \author{Name\corauthref{cor1}\thanksref{label2}}
% \ead{email address}
% \ead[url]{home page}
% \thanks[label2]{}
% \corauth[cor1]{}
% \address{Address\thanksref{label3}}
% \thanks[label3]{}

\title{On Radiation Emission from a Microbunched Beam with Wavefront Tilt and Its Experimental Observation}

% use optional labels to link authors explicitly to addresses:
% \author[label1,label2]{}
% \address[label1]{}
% \address[label2]{}

\author[XFEL]{Gianluca Geloni,}
\author[DESY]{Vitali Kocharyan,}
\author[DESY]{Evgeni Saldin}
\address[XFEL]{European XFEL GmbH, Hamburg, Germany}
\address[DESY]{Deutsches Elektronen-Synchrotron (DESY), Hamburg, Germany}

\begin{abstract}
In this paper we compare experimental observations and theory of radiation emission from a microbunched beam with microbunching wavefront tilt with respect to the direction of motion. The theory refers to the work  \cite{KITA}, which predicts, in this case, exponential suppression of coherent radiation along the kicked direction.  The observations refer to a recent experiment performed at the LCLS \cite{NUHN,NATAL}, where a microbunched beam was kicked by a bend and sent to a radiator undulator. The experiment resulted in the emission of strong coherent radiation that had its maximum along the kicked direction of motion, when the undulator parameter was detuned to a value  larger than the nominal one.  We first analyze the theory in detail, and we confirm the correctness of its derivation according to the conventional theory of radiation emission from charged particles. Subsequently, we look for possible peculiarities in the experiment, which may not be modeled by the theory. We show that only spurious effects are not accounted for. We conclude that the experiment defies explanation in terms of the conventional theory of radiation emission.
\end{abstract}

%\begin{keyword}
%
%% keywords here, in the form: keyword \sep keyword
%%edge radiation \sep near-field \sep electron-bunch diagnostics
%%\sep x-ray free-electron laser (XFEL)
%

%% PACS codes here, in the form: \PACS code \sep code
%%\PACS 41.60.Cr \sep 42.25.-p \sep 41.75.-Ht
%\end{keyword}
%
\end{frontmatter}

% main text

%\linenumbers

\section{Introduction and motivation}

The theory of spontaneous emission from a microbunched beam with wavefront tilt with respect to the velocity of propagation has been developed several years ago \cite{KITA}. When the microbunching wavefront tilt becomes larger than the coherence angle, this theory predicts a dramatic suppression of coherent emission in the beam propagation direction. In relation to this result, a recent experiment at the LCLS \cite{NUHN,NATAL} yielded remarkable outcomes.

The LCLS generates linearly polarized X-ray pulses from a planar undulator. A $3.2$ m-long Delta undulator, which allows for a full control of the degree of polarization of the emitted radiation, was recently installed in place of the last LCLS undulator segment.  Before going through the  Delta undulator, the electron beam is microbunched in the preceding planar undulator segments. This enhances the radiation power by several orders of magnitude. Therefore, the Delta undulator is said to be operating in `afterburner configuration'. Such configuration leads to the presence of linearly-polarized background radiation from the main undulator, which should be suppressed. In fact, when the efficiency of the regular afterburner mode of operation was tested, a maximum contrast ratio of about $2.5$ was achieved \cite{NUHN}. It has been recently proposed \cite{REVE} that the background radiation component can be greatly reduced by a reverse undulator tapering configuration. By inverting the sign of the baseline undulator tapering the radiation emission is reduced, while microbunching can still develop. The efficiency of this mode of operation was tested and a contrast ratio of about $10$ was reported in \cite{NUHN}. From a practical viewpoint, under this conditions, at the entrance of the Delta undulator there is only a micro-bunched beam. References \cite{NUHN,NATAL} further report a final improvement of the degree of polarization  up to $100 \%$ by X-ray beam splitting at the photon energy of $0.7$ keV. This was achieved by kicking the electron beam before entering the Delta undulator, in order to let electron beam and background radiation pass through the Delta undulator at different angles. The quadrupole at the end of the last planar undulator section includes a regular vertical corrector, which was used to control the magnitude of the kick. According to \cite{NUHN}, the maximal kick angle was about $3 \times 10^{-5}$ rad and was limited only by the $4$ mm diameter of the beamline aperture at the distance of 80 m. At this maximum angle, the separation between the two radiation spots on the screen in the experimental hall was about $5$ rms times the radiation spot size. Moreover, the energy of the output radiation pulse with and without kick is practically the same.

\begin{figure}
\centering
\includegraphics[width=0.6\textwidth]{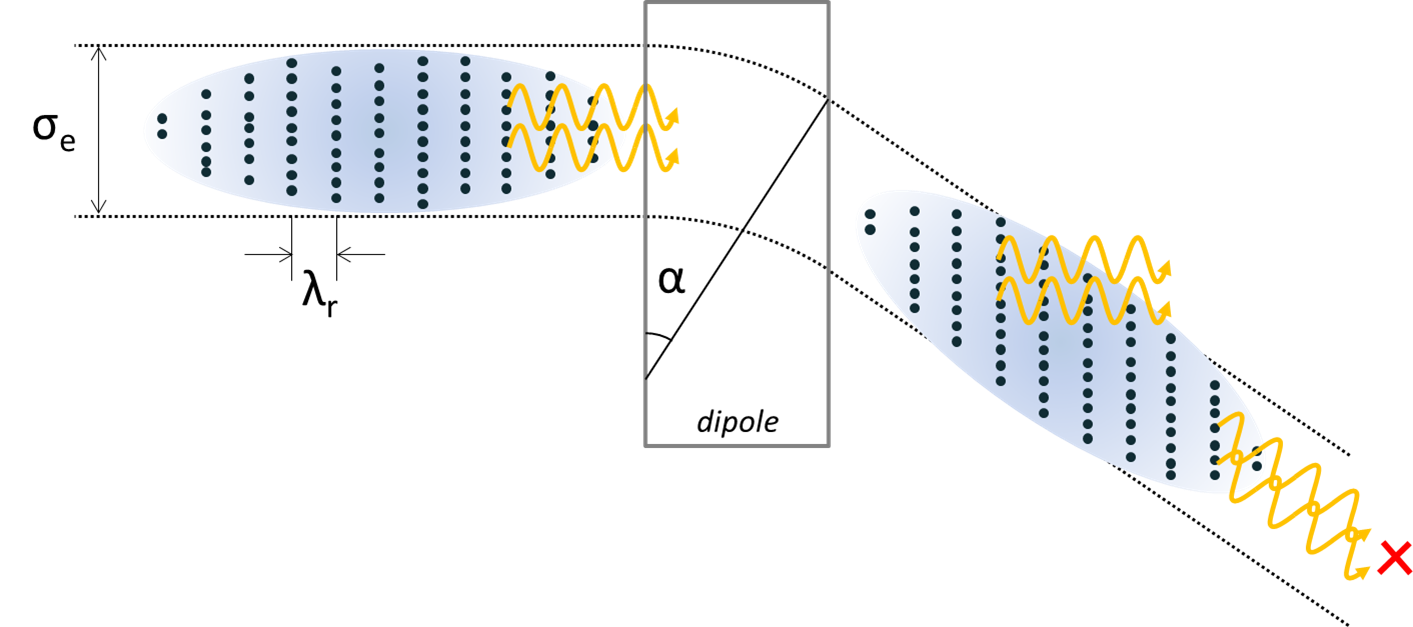}% \hfil% Here is how to import EPS art
\caption{Illustration of the problem, which arises according to classical particle tracking when a microbunched electron beam is deflected by a dipole magnet. After passing the dipole, the microbunching is preserved, but only along its original direction.}
\label{classical}
\end{figure}

In order to explain this observation in relation with the theory in \cite{KITA}, one would conclude that the LCLS experiment apparently shows a readjustment of the microbunching orientation in the kicked direction. In this way, one could produce coherent radiation in the kicked direction. However, classical particle tracking shows that while the electron beam direction changes after the kick, the orientation of the microbunching wavefront stays unvaried. Therefore, the electron motion and the wavefront normal have different directions. Figure \ref{classical} illustrates the issue. If one assumes that no readjustment of the microbunching wavefront takes place, according to \cite{KITA} the FEL process in the downstream undulator is expected to be dramatically suppressed because the kick angle is larger than the divergence of the output coherent radiation, in contrast with the experimental observation.

In order to estimate the loss of radiation efficiency we make the assumption that the spatial profile of the bunching factor is close to that of the electron beam and has a Gaussian shape with standard deviation $\sigma_\mathrm{b}$. A bunched electron beam in an FEL amplifier can be considered as a sequence of periodically spaced oscillators. The radiation produced by these oscillators always interferes coherently at zero angle with respect to the undulator axis. In the limit for a small size of the electron beam the interference will be constructive within an angle of about $\Delta\theta_\mathrm{c} \backsimeq \sqrt{c/(\omega L_\mathrm{g})}$, where $L_\mathrm{g}$ is the FEL gain length. In the limit for a large size of the electron beam, the angle of coherence is about $\Delta\theta_\mathrm{c} \backsimeq c/(\omega\sigma_\mathrm{b})$ instead. The boundary between these two asymptotes is for sizes of about $\sigma_\mathrm{dif} \backsimeq \sqrt{cL_\mathrm{g}/\omega}$. It is worth noting that the condition $\sigma_\mathrm{b}^2 \gg \sigma_\mathrm{dif}^2$ is satisfied in our case of study at the LCLS. Thus, we can conclude that the angular distribution of the radiation power in the far zone has a Gaussian shape with standard deviation $\sigma_\mathrm{c} \backsimeq c/(\sqrt{2}\omega\sigma_\mathrm{b})$. After the electron beam is kicked, as already mentioned, in classical treatments we have a discrepancy between direction of the electron motion and wavefront normal. Then, the radiation intensity along the new direction of the electron beam can be approximated as $I \backsimeq I_0  \exp[- \theta^2/(2\sigma_\mathrm{c}^2)]$, where $I_0$ is the  on-axis intensity without kick and $\theta$ is the kick angle.  The exponential suppression factor is due to the microbunching wavefront tilt with respect to the direction of motion of the electrons. Beam splitting at the LCLS was done by kicking the electron beam of an angle of about $5$ standard deviations of the intensity distribution in the far zone. According to the estimations presented above, the intensity of the coherent radiation in the kicked direction should be suppressed by two orders of magnitude. In spite of this, the experiment showed that the radiation intensity  in the kicked direction is practically the same as the intensity without kick at zero angle.

In addition to exponential suppression of the intensity, one expects negligible detuning effects in the case of radiation emitted along the direction of the kick. In fact, the effective undulator period is now given by $\lambda_{w}/\cos(\theta) \backsimeq (1 + \theta^2/2)\lambda_{w}$, where $\lambda_{w}$ is the actual undulator period. This induces a relative red shift in the resonance wavelength of about $\Delta\lambda/\lambda \backsimeq \theta^2/2$ which should be compared with the relative bandwidth of the resonance, the $\rho$ parameter, which is much larger. As a result, the red shift in the resonance wavelength due to the kick can be neglected in all situations of practical relevance. It is clear from the above that if a microbunched beam is at perfect resonance along the direction of motion without kick, then after the kick the same microbunched beam is at perfect resonance along the new direction of the electron beam motion. However, references \cite{NUHN,NATAL} report that the radiation in the kicked direction is red-shifted with respect to the case when no kick is applied.

This experimental result is in contradiction with the theory \cite{KITA}. In this paper we analyze such contradiction in detail. Logically speaking there are three possibilities to explain this contradiction. First, the theory in \cite{KITA} is, for some reason, incorrect. Second, the theory in \cite{KITA} does not model the actual experiment, i.e. there are some peculiarities of the experiment that are not accounted for in the theory. In this case, such peculiarity could be, for example, a readjustment of the microbunching wavefront, which is not foreseen according to usual particle tracking. Third, there are no peculiarities in the experiment that are neglected in the theory in \cite{KITA}, and the theory is correctly derived on the basis of the usual laws of electrodynamics and dynamics of charged particles, so there must be some more fundamental reason for the discrepancy observed.

In the next Section we critically review the theory in \cite{KITA} and we show that it is correctly derived according to the usual laws of electrodynamics and dynamics of charged particles. Subsequently we discuss possible reasons why the experiment might not be fully modeled by this theory. Our conclusion is that there are none: in our view, a readjustment in the microbunching wavefront due to neglected dynamical effects is to be excluded. We argue that, in order to explain the reason why the experiment in \cite{NUHN,NATAL} is in disagreement with the result by \cite{KITA}, more fundamental reasons should be invoked.

\section{Undulator radiation from a microbunched electron beam}

Let us critically review the theory in \cite{KITA}. We consider an electron beam modulated in density at a single frequency $\omega$ as source of coherent undulator radiation. We write the longitudinal current density $j_z$ as a sum of two terms: a constant unperturbed term, $j_{o}$, and a term describing the actual modulation at frequency $\omega$ $\tilde{j}_{z}$, to be considered as a perturbation:

\begin{equation}
j_z(z,\vec{r}_\bot,t) = j_{o}(z,\vec{r}_\bot) + \tilde{j}_{z}(z,\vec{r}_\bot,t)~, \label{sum1}
\end{equation}
with $\vec{r}_\bot$ a two-dimensional vector fixing the transverse coordinates and $t$ the time. We follow \cite{2HAR} write the unperturbed part $j_{o}$ as as

\begin{equation}
j_{o}(z,\vec{r}_\bot) = {j}_{o}(\vec{r}_\bot - \vec{r}^{(c)}_\bot(z))~, \label{unp}
\end{equation}
where $\vec{r}^{(c)}_\bot(z)$ describes a "coherent motion" followed by all particles. In the special case of a single electron $j_o$ is  a $\delta$-Dirac function. Eq. (\ref{unp}) is certainly valid in the case of a monochromatic beam, assuming homogeneous undulator field in the transverse direction.  Eq. (\ref{unp}) is also valid in the case of finite energy spread if the transverse size of the electron beam is larger than the transverse excursion of the electrons during their wiggling motion: the validity of Eq. (\ref{unp}) has an accuracy given by the relative deviation of the particles energy form the average value, $\delta \gamma/\gamma$ -with $\gamma$ the usual Lorentz factor, and must be small if the electron beam is used for Free-Electron Laser light generation.

Following the same notation in \cite{2HAR} we write

\begin{eqnarray}
\tilde{j}_{z}(z,t) &=&
j_{o}\left(\vec{r}_\bot-\vec{r}^{(c)}_\bot(z)\right) \cr &&\times
\left\{\tilde{a}_1\left(z,\vec{r}_\bot-\vec{r}^{(c)}_\bot(z)\right)\exp\left[i
\omega \int_0^z \frac{dz'}{v_z(z')} - i \omega t
\right]+\mathrm{C.C.}\right\}~, \label{jzp}
\end{eqnarray}
where "C.C." indicates the complex conjugate of the first term in parenthesis. Here $\tilde{a}_1$ is to be considered as a given complex function describing the evolution of the microbunching, while the longitudinal velocity ${v}_z(z)$ can be recovered from the knowledge of $\vec{r}^{(c)}_\bot(z)$ and of the average energy of the beam $\gamma=\gamma(z)$.

We introduce the possibility of beam deflection angles $\eta_x$ (horizontal) and $\eta_y$ (vertical) with respect to the $z$ axis and we indicate the motion in absence of deflection with the subscript "(nd)". In the case of a short undulator with no focusing elements in between, one simply obtains:

\begin{eqnarray}
v_z(z,\eta) &=& v_{z(nd)}(z)
\left(1-\frac{\eta_x^2+\eta_y^2}{2}\right) \cr
\vec{v}_\bot(z,\eta) &=& \vec{v}_{\bot (nd)}(z) + v_{z(nd)}(z)
\vec{\eta}~, ~~~~\label{etav}
\end{eqnarray}
and

\begin{eqnarray}
\vec{r}^{(c)}_\bot(z,\vec{\eta}) = \vec{r}^{(c)}_{\bot(nd)}(z) +
\vec{\eta} z ~. ~~~~\label{etar}
\end{eqnarray}
The orientation of the microbunching wavefront has an impact on the way $\tilde{a}_1$ depends on $\vec{\eta}$ and can be kept fully general at this stage setting

\begin{equation}
\tilde{a}_1 =
\tilde{a}_1\left(z,\vec{r}_\bot-\vec{r}^{(c)}_\bot(z,\vec{\eta})\right)
~.\label{roto}
\end{equation}
Still following \cite{2HAR}, in the limit for $\gamma^2 \gg 1$, the total current density can be written as

\begin{eqnarray}
\vec{j}(z,t,\vec{\eta})&=&\frac{\vec{v}(z,\vec{\eta})}{c}
j_{o}\left(\vec{r}_\bot-\vec{r}^{(c)}_\bot(z,\vec{\eta})\right)
\Bigg\{1+\Bigg[\tilde{a}_1\left(z,
\vec{r}_\bot-\vec{r}^{(c)}_\bot(z,\vec{\eta})\right)\cr&&\times \exp\Bigg[i
\omega \int_0^z \frac{dz'}{v_z(z',\vec{\eta})}-i\omega t
\Bigg]+\mathrm{C.C.}\Bigg]\Bigg\}~, \label{totcur}
\end{eqnarray}
where $c$ is the speed of light in vacuulm while the charge density is

\begin{equation}
\rho =\frac{j_z}{v_z} \simeq \frac{j_z}{c}~, \label{chd}
\end{equation}
since we work under the paraxial approximation.

We look for solutions of the inhomogeneous wave equation for the electric field $\vec{E}_\bot$ in the form

\begin{equation}
\vec{{E}}_\bot = \vec{\widetilde{E}}_\bot \exp\left[i\omega
(z/c-t)\right] + \mathrm{C.C.}~\label{etildadef}
\end{equation}
in the case of undulator emission. If the electric field does not vary much over an undulator period, $\vec{\widetilde{E}}_\bot$ has the physical meaning of a slowly varying envelope. The wave equation in paraxial approximation can then be written (see \cite{2HAR}) as

\begin{eqnarray}
\left({\nabla_\bot}^2 + \frac{2 i \omega}{c}
\frac{\partial}{\partial z}\right) \vec{\widetilde{E}}_{\bot} =
\frac{4 \pi}{c}  \exp\left[i \left(\Phi_s-\omega
\frac{z}{c}\right)\right] \left[\frac{i\omega}{c^2}\vec{v}_\bot
-\vec{\nabla}_\bot \right] j_o \tilde{a}_1 ~. \label{incipit4}
\end{eqnarray}

where

\begin{equation}
\Phi_s(z,\vec{\eta}) = \omega \int_0^{z} \frac{d
z'}{v_z(z',\vec{\eta})} ~, \label{phu}
\end{equation}
We want to solve Eq. (\ref{incipit4})  for a planar undulator where electrons have velocities

\begin{figure}
	\centering
	\includegraphics[width=0.8\textwidth]{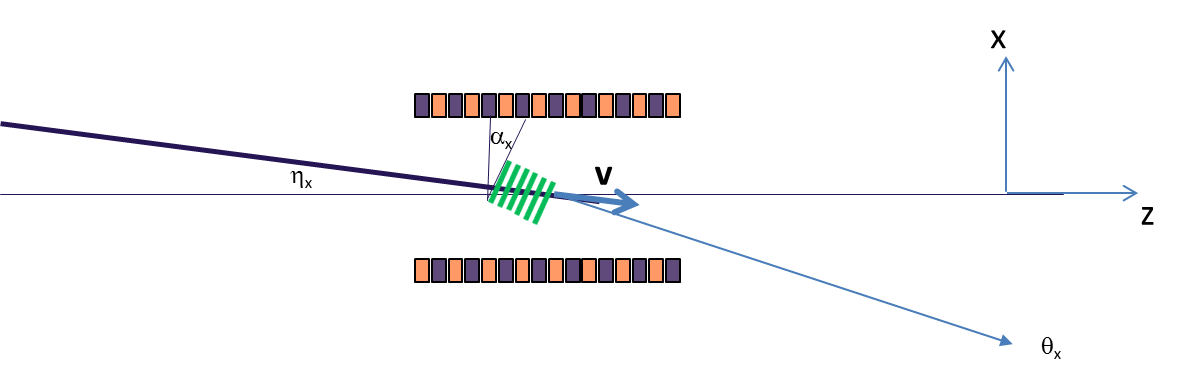}% \hfil%
	\caption{Geometry of the problem. The angles $\vec{\alpha}$  control the tilt of the microbunching wavefront with respect to the $z$ direction, while the angles $\vec{\eta}$ control the direction of the beam with respect to the $z$ direction. When $\vec{\alpha} = 0$ the normal to the microbunching wavefront is along $z$, while when $\vec{\eta} = 0$ the velocity of the beam is along $z$. One observes radiation at angles $\vec{\theta}$ with respect to the $z$ axis.}
	\label{figure0}
\end{figure}

\begin{figure}
	\centering
	\includegraphics[width=0.8\textwidth]{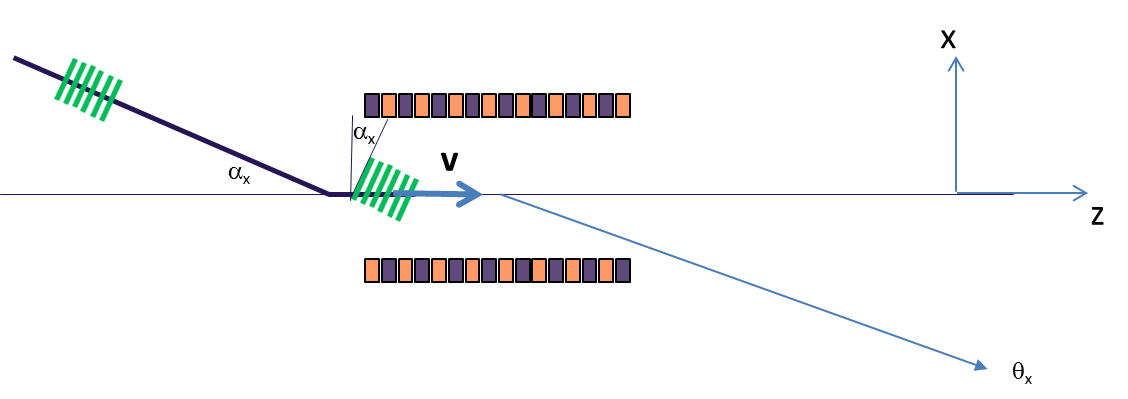}% \hfil%
	\caption{Geometry of the problem, simplified case for $\vec{\alpha}=(\alpha_x,0)$ and $\vec{\eta} = 0$. This can be interpreted as the case where a microbunched beam is kicked by an angle $\alpha$ and enters an undulator radiator along the direction of the $z$ axis.}
	\label{figure1}
\end{figure}

\begin{equation}
\vec{v}_\bot(z',\vec{\eta}) = \left[- {c K\over{\gamma}}
\sin{\left(k_w z'\right)}+\eta_x v_z\right] \vec{x}+\left[\eta_y
v_z \right]\vec{y}~, \label{vuz2}
\end{equation}
and follow the constrained motion

\begin{eqnarray}
\vec{r}^{(c)}_\bot(z',\vec{\eta})+\vec{l} = \left[ \frac{K}{\gamma
k_w} \left(\cos{\left(k_w z'\right)}-1\right)+\eta_x z'
+l_x\right] \vec{x}  + \left[\eta_y z' +l_y\right]\vec{y}~.
\label{erz2}
\end{eqnarray}
Here $l_x$ and $l_y$ model the electron beam offset, while $k_w = 2\pi/\lambda_w$, $\lambda_w$ being the undulator period. We define the undulator parameter $K$ as

\begin{equation}
K=\frac{\lambda_w e H_w}{2 \pi m_\mathrm{e} c^2}~, \label{Kpara}
\end{equation}
with $(-e)$ is the (negative) electron charge, $m_\mathrm{e}$  the electron mass, and $H_w$ is the maximum of the magnetic field produced by the undulator on the $z$ axis.

Eq. (\ref{incipit4}) can be solved by means of a proper Green's function choice. Still following \cite{2HAR} we have

\begin{eqnarray}
\widetilde{\vec{E}}_{\bot }(z_o, \vec{r}_{\bot o} )&=&
-\frac{1}{c}\int_{-\infty}^{\infty} dz' \frac{1}{z_o-z'} \int d
\vec{r'}_{\bot}
\left[\frac{i\omega}{c^2}\vec{v}_\bot(z',\vec{\eta})
-\vec{\nabla}'_\bot \right]\cr &&\times
j_{o}\left(\vec{r'}_\bot-\vec{r}^{(c)}_\bot(z',\vec{\eta})\right)
\tilde{a}_2 \left(z',\vec{r'}_\bot
-\vec{r}^{(c)}_\bot(z',\vec{\eta}) \right)\cr&&
\exp\left\{i\omega\left[\frac{\mid \vec{r}_{\bot o}-\vec{r'}_\bot
\mid^2}{2c (z_o-z')}\right]+ i \left[ \Phi_s(z',\vec{\eta})-\omega
\frac{z'}{c}\right] \right\} ~, \label{blob}
\end{eqnarray}
$\vec{\nabla}'_\bot$ being the gradient operator with respect to the source point. Moreover, $(z_o, \vec{r}_{\bot o})$ is the observation point. Further integration by parts of the gradient terms gives

\begin{eqnarray}
\widetilde{\vec{E}}_{\bot}&= &-\frac{i \omega }{c^2}
\int_{-\infty}^{\infty} dz' \frac{1}{z_o-z'}  \int d
\vec{r'}_{\bot} \left(\frac{\vec{v}_\bot(z',\vec{\eta})}{c}
-\frac{\vec{r}_{\bot o}-\vec{r'}_\bot}{z_o-z'}\right)\cr&&\times
j_{o}\left(\vec{r'}_\bot-\vec{r}^{(c)}_\bot(z',\vec{\eta})\right)
\tilde{a}_2 \left(z',\vec{r'}_\bot
-\vec{r}^{(c)}_\bot(z',\vec{\eta}) \right) \exp\left[i
\Phi_T(z',\vec{r'}_\bot,\vec{\eta})\right] ~, \cr &&
\label{generalfin}
\end{eqnarray}
where the total phase $\Phi_T$ is given by

\begin{equation}
\Phi_T =  \left[\Phi_s-\omega\frac{z'}{c}\right]+ \omega \left[
\frac{|\vec{r}_{\bot o}-\vec{r'}_\bot|^2}{2c (z_o-z')}\right]~.
\label{totph}
\end{equation}
We will keep only resonant terms and look near the first harmonic in the far zone. Moreover, we will make use of a new integration variable $\vec{l'}=\vec{r'}_\bot-\vec{r}^{(c)}_\bot(z',\vec{\eta})$. We omit detailed calculations: the interested reader may follow a detailed derivation for the second harmonic in \cite{2HAR}). The overall result is

\begin{eqnarray}
\vec{\widetilde{E}}_{\bot}&&=- \frac{K \omega A_{JJ}}{2 \gamma c^2 z_o} \exp\left[\frac{i \omega z_0 \theta^2}{2c}\right]
\int d \vec{l}'\int_{-L_w/2}^{L_w/2} d z' \cr && \times \exp\left[-\frac{i \omega}{c}\vec{\theta}\cdot\vec{l}'\right]\exp\left[\frac{i \omega z'}{2c} \left(\vec{\theta}-\vec{\eta}\right)^2\right]\exp\left[i z' C\right] \tilde{\rho}^{(1)}(z',\vec{l},C)~,
\label{undurad4}
\end{eqnarray}
where we have defined

\begin{equation}
\tilde{\rho}^{(1)}({z}',\vec{l},{C}) = j_{o}\left(\vec{l}\right)
\tilde{a}_1 \left(z',\vec{l} \right)  ~,\label{rhodefin}
\end{equation}
In Eq. (\ref{undurad4}) $\vec{\theta}$ is the observation angle, $A_{JJ} = J_0[K^2/(4+2K^2)] - J_1[K^2/(4+2K^2)]$ and the detuning from resonance is

\begin{equation}
C = \frac{\omega - \omega_{1}}{\omega_{1}} k_w \label{Cdef}~.
\end{equation}
where the resonance frequency $\omega_1$ is

\begin{equation}
\omega_{1} = 2 {k_w c {\gamma}_z^2} ~.\label{freqfix10}
\end{equation}
If we consider $\tilde{\rho}^{(1)}$ as a given function we can allow for any particular presentation of the beam modulation. We now introduce a model for $\tilde{\rho}^{(1)}$:

We now consider the case when  $\gamma(z) = \bar{\gamma}= \mathrm{const}$,  and when

\begin{equation}
\tilde{\rho}^{(1)}({z},\vec{l}) = j_{o}\left(\vec{l}\right) a_{1}
\exp \left[i \frac{\omega_{1}}{c} \vec{\alpha} \cdot \vec{l} \right] ~, \label{expara}
\end{equation}
with $a_{1} = \mathrm{const}$  and

\begin{equation}
{j}_{o}\left(\vec{{l}}\right) = \frac{I_o}{2\pi {\sigma^2}}
\exp{\left(-\frac{{l}_x^2+{l}_y^2}{2 \sigma^2}\right)}~.
\label{exbot}
\end{equation}
Here $I_o$ and $\sigma$ are the bunch current and transverse size respectively.

\begin{figure}
	\centering
	\includegraphics[width=0.8\textwidth]{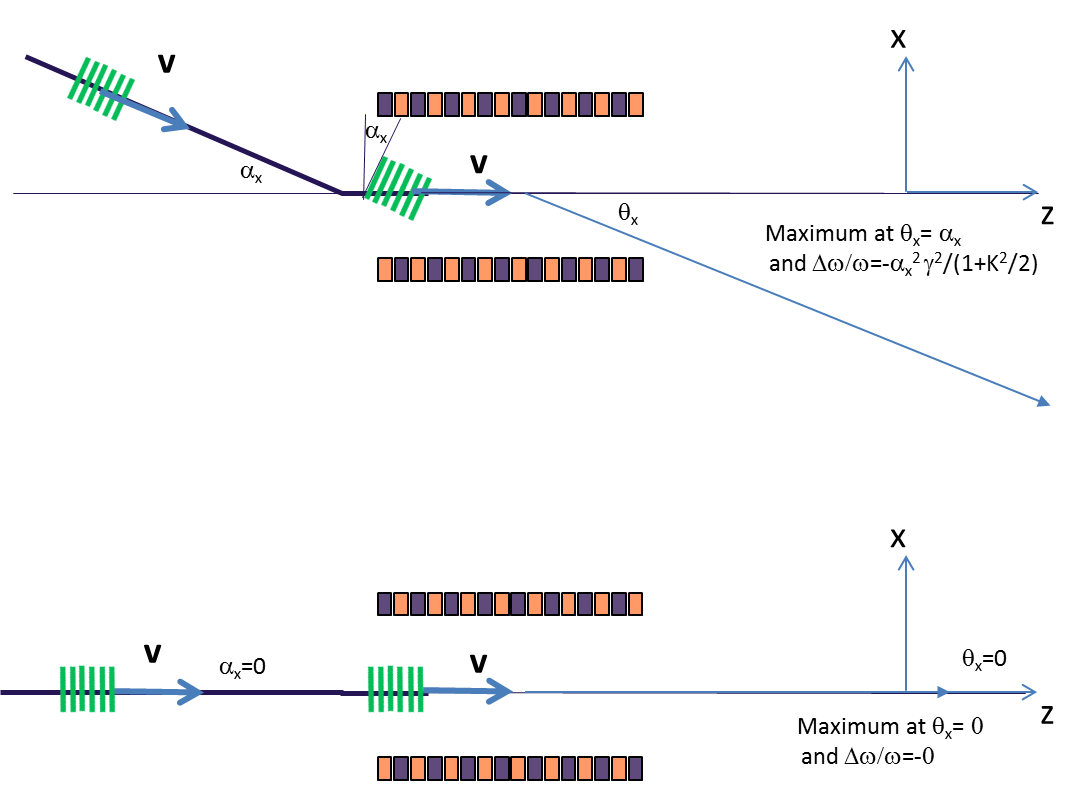}% \hfil%
	\caption{Maximum emission comparison for microbunched beam entering the undulator after a kick or without a kick, according to conventional theory}
	\label{figure2}
\end{figure}

Substitution of Eq. (\ref{expara}) into Eq. (\ref{undurad4}) and integration yields

\begin{eqnarray}
\widetilde{E}_{\bot}&&=- \frac{K \omega A_{JJ} I_o L_w a_1}{2 \gamma c^2 z_o} \exp\left[\frac{i \omega z_0 \theta^2}{2c}\right] \mathrm{sinc}\left[\frac{L_w}{2} \left(C+\frac{\omega \left| \vec{\theta} - \vec{\eta}\right|^2}{2c}\right)\right] \cr && \times \exp\left[-\frac{\sigma^2 \omega^2}{2 c^2}\left| \vec{\theta} - \vec{\alpha}\right|^2\right]~.
\label{unduradoutgen}
\end{eqnarray}
that is a scalar because in the resonant approximation the field is horizonatlly polarized. The angles $\vec{\alpha}$  control the tilt of the microbunching wavefront with respect to the $z$ direction, while the angles $\vec{\eta}$ control the direction of the beam with respect to the $z$ direction. When $\vec{\alpha} = 0$ the normal to the microbunching wavefront is along $z$, while when $\vec{\eta} = 0$ the velocity of the beam is along $z$. One observes radiation at angles $\vec{\theta}$ with respect to the $z$ axis. The overall geometry is summarized in Fig. \ref{figure0}.

The associated power is given  by

\begin{eqnarray}
W = \frac{c}{4 \pi} \int_{-\infty}^{\infty} dx_o
\int_{-\infty}^{\infty} dy_o \overline{|E_{\bot}(z_o, x_o,
y_o,t)|^2} &&\cr= \frac{c}{2 \pi} \int_{-\infty}^{\infty} dx_o
\int_{-\infty}^{\infty} dy_o {|\tilde{E}_{\bot}(z_o, x_o,
y_o)|^2} = \frac{c}{2 \pi}  \left(\frac{K \omega A_{JJ} I_o L_w a_1}{2 \gamma c^2 }\right)^2&&\cr\times \int d\vec{\theta}~ \mathrm{sinc}^2\left[\frac{L_w}{2} \left(C+\frac{\omega \left| \vec{\theta} - \vec{\eta}\right|^2}{2c}\right)\right]  \exp\left[-\frac{\sigma^2 \omega^2}{ c^2}\left| \vec{\theta} - \vec{\alpha}\right|^2\right]\label{xpowden}
\end{eqnarray}
where $\overline{(...)}$ denotes averaging over a cycle of oscillation of the carrier wave.

Note that introducing the angle $\vec{\xi} = \vec{\theta} - \vec{\alpha}$ between the observation direction and the normal to the microbunching, one can cast the power in the form of a convolution between the single-particle emission and the Fourier transform of the transverse electron beam distribution

\begin{eqnarray}
W \propto \int d\vec{\xi}~ \mathrm{sinc}^2\left[\frac{L_w}{2} \left(C+\frac{\omega \left| \vec{\xi} - (\vec{\eta}-\vec{\alpha})\right|^2}{2c}\right)\right]  \exp\left[-\frac{\sigma^2 \omega^2}{ c^2}\xi^2\right]~.\label{xpowden2}
\end{eqnarray}
One thus recovers Eq. (4) of reference \cite{KITA}, where the power is shown to be a function of the angle between the electron beam direction and the microbunching wavefront normal, that is $\vec{\eta}-\vec{\alpha}$ according to our notations.

For further analysis we introduce normalized units for all angular quantities $\phi$ as $\hat{\phi} = \phi/[c/(\omega L_w)] \phi$, with the physical meaning of angles normalized to the diffraction angle, for the distance $\hat{z}_o = z/L_w$, for the detuning from resonance $\hat{C} = C L_w = 2 \pi N_w (\omega-\omega_1)/\omega$, for the electron beam size (basically a Fresnel number) $\hat{N} = \omega \sigma^2/(L_w c)$, for the electric field $\hat{E} = \widetilde{E}_{\bot} K \omega A_{JJ} I_o L_w a_1 /(2 \gamma c^2 z_o)$, for the intensity $\hat{I} = |\hat{E}|^2$, and for the power $\hat{W} = \int d\vec{\hat{\theta}} \hat{I}$. These definitions give

\begin{eqnarray}
\hat{E} &&=\exp\left[\frac{i \hat{\theta}^2 \hat{z}_o}{2}\right] \mathrm{sinc}\left[\frac{\hat{C}}{2}+\frac{\left| \vec{\hat{\theta}} - \vec{\hat{\eta}}\right|^2}{4}\right]  \exp\left[-\frac{\hat{N}}{2}\left| \vec{\hat{\theta}} - \vec{\hat{\alpha}}\right|^2\right]~.
\label{unduradoutgennorm}
\end{eqnarray}
\begin{eqnarray}
\hat{I} &&= \mathrm{sinc}^2\left[\frac{\hat{C}}{2}+\frac{\left| \vec{\hat{\theta}} - \vec{\hat{\eta}}\right|^2}{4}\right]  \exp\left[-\hat{N}\left| \vec{\hat{\theta}} - \vec{\hat{\alpha}}\right|^2\right]~.
\label{unduradintenorm}
\end{eqnarray}
and

\begin{eqnarray}
\hat{W} &&=  \int d\vec{\hat{\xi}}~ \mathrm{sinc}^2\left[\frac{\hat{C}}{2}+\frac{\left| \vec{\hat{\xi}} - (\vec{\hat{\eta}}-\vec{\hat{\alpha}})\right|^2}{4}\right]  \exp\left[-\hat{N}\xi^2\right]~.
\label{unduradWnorm}
\end{eqnarray}
In order to simplify the study case without losing in generality we assume  $\vec{\alpha}=(\alpha_x,0)$ and $\vec{\eta} = 0$. This can be interpreted as the case depicted in Fig. \ref{figure1}, where a microbunched beam is kicked by an angle $\alpha$ and enters an undulator radiator along the direction of the $z$ axis. Then, looking at $\theta_y=0$ we have the following profile for $\hat{I}$:

\begin{eqnarray}
\hat{I} &&= \mathrm{sinc}^2\left[\frac{\hat{C}}{2}+\frac{ \hat{\theta}_x^2}{4}\right]  \exp\left[-{\hat{N}}( {\hat{\theta}_x} - {\hat{\alpha}_x})^2\right]~.
\label{unduradintenorm2}
\end{eqnarray}
while from Eq. (\ref{unduradWnorm}) we obtain:

\begin{eqnarray}
\hat{W} &&=   \int d \hat{\xi}_x \int d\hat{\xi}_y ~ \mathrm{sinc}^2\left[\frac{\hat{C}}{2}+\frac{\hat{\xi}_x^2}{4}+\frac{\hat{\xi}_y^2}{4}\right]  \exp\left[-\hat{N}(\hat{\xi}_x-\hat{\alpha}_x)^2\right]\exp\left[-\hat{N} \hat{\xi}_y^2\right]~.\cr &&
\label{unduradWnorm2}
\end{eqnarray}
Note that while we performed our calculations for the case of a planar undulator, the dimensionless result in Eq. (\ref{unduradintenorm2}) remains valid also for the helical case. Inspection of Eq. (\ref{unduradintenorm2}) shows that the radiation maximum is for $\hat{\theta}_x = \hat{\alpha}_x$ and red-shifted of $\hat{C} = \hat{\alpha}^2/2$ or, in dimensional units, $\theta_x = \alpha_x$ and $C = - \omega \alpha_x^2 /(2 c) = - k_w \alpha_x^2 \gamma^2/(1 +K^2/2)$. Summing up, for $\alpha_x \ne 0$, the conventional theory predicts emission of radiation as in Fig. \ref{figure2}(top) compared to the case for $\alpha_x = 0$ in Fig. \ref{figure2}(bottom).

We then fix $\hat{C} = \hat{\alpha}_x^2/2$ and we illustrate further some asymptotic behaviors. First, Eq. (\ref{unduradintenorm2}) and Eq. (\ref{unduradWnorm2}) that become

\begin{eqnarray}
\hat{I} &&= \mathrm{sinc}^2\left[-\frac{\hat{\alpha}_x^2}{4}+\frac{ \hat{\theta}_x^2}{4}\right]  \exp\left[-\hat{N}( {\hat{\theta}_x} - {\hat{\alpha}_x})^2\right]~.
\label{unduradintenorm3}
\end{eqnarray}
\begin{eqnarray}
\hat{W} &&=   \int d \hat{\xi}_x \int d\hat{\xi}_y ~ \mathrm{sinc}^2\left[-\frac{\hat{\alpha}_x^2}{4}+\frac{\hat{\xi}_x^2}{4}+\frac{\hat{\xi}_y^2}{4}\right]  \exp\left[-\hat{N}(\hat{\xi}_x-\hat{\alpha}_x)^2\right]\exp\left[-\hat{N} \hat{\xi}_y^2\right]~.\cr &&
\label{unduradWnorm3}
\end{eqnarray}
\begin{figure}
	\centering
	\includegraphics[width=0.45\textwidth]{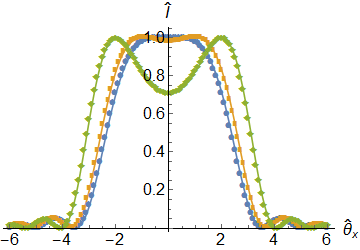}% \hfil%
    \includegraphics[width=0.45\textwidth]{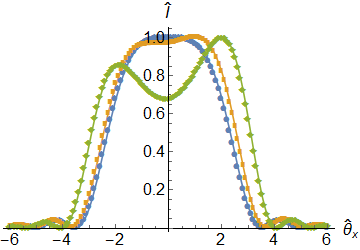}
    \includegraphics[width=0.45\textwidth]{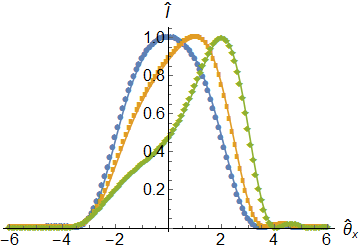}
    \includegraphics[width=0.45\textwidth]{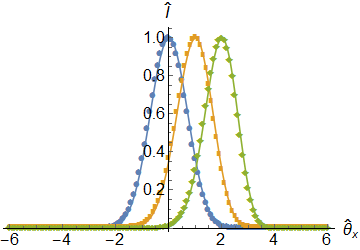}
    \includegraphics[width=0.45\textwidth]{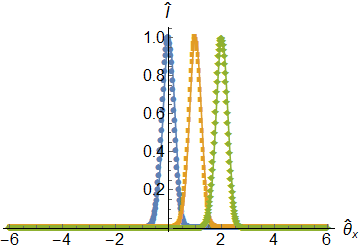}
    \includegraphics[width=0.45\textwidth]{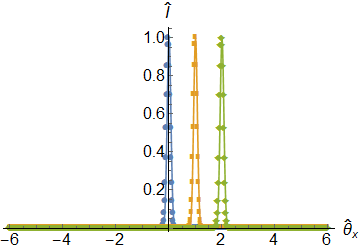}
	\caption{From left to right and top to bottom: the behavior of $\hat{I}$ in Eq. (\ref{unduradintenorm3}) for several values of $\hat{N}=10^{-4}$, $\hat{N}=10^{-2}$, $\hat{N}=10^{-1}$, $\hat{N}=1$, $\hat{N}=10$, $\hat{N}=100$ and various $\hat{\alpha}_x = 0,1,2$ (blue circles, orange square, green diamonds). }
	\label{figure3}
\end{figure}
We show the behavior of $\hat{I}$ in Eq. (\ref{unduradintenorm3}) for several values of $\hat{N}$ and various $\hat{\alpha}_x$ in Fig. \ref{figure3}.

When $\hat{\alpha}_x \ll 1$ and $\hat{N} \hat{\alpha}_x \ll 1$ one finds the well-known limiting relations for $\hat{\alpha}_x=0$ (see e.g. \cite{SSY1})

\begin{eqnarray}
\hat{I} &&\equiv \hat{I}_0(\hat{N},\hat{\theta}_x) =  \mathrm{sinc}^2\left[\frac{ \hat{\theta}_x^2}{4}\right]  \exp\left[-\hat{N} \hat{\theta}_x^2\right]~.
\label{unduradintenormlim1a}
\end{eqnarray}
\begin{eqnarray}
\hat{W} && \equiv  \hat{W}_0(\hat{N}) = \int d \hat{\xi}_x \int d\hat{\xi}_y ~ \mathrm{sinc}^2\left[\frac{\hat{\xi}_x^2}{4}+\frac{\hat{\xi}_y^2}{4}\right]  \exp\left[-\hat{N}(\hat{\xi}_x^2+ \hat{\xi}_y^2)\right]~.\cr && = 4\pi \left[ \arctan\left(\frac{1}{2\hat{N}}\right) +N \ln\left(\frac{4 \hat{N}^2}{4 \hat{N}^2 +1}\right)\right]
\label{unduradWnormlim1a}
\end{eqnarray}
\begin{figure}
	\centering
	\includegraphics[width=0.5\textwidth]{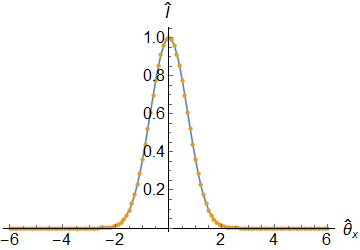}% \hfil%
    \includegraphics[width=0.5\textwidth]{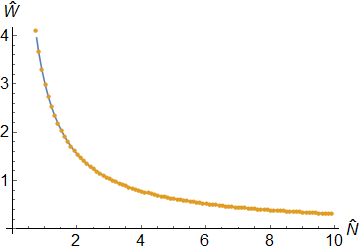}% \hfil%
	\caption{Left: $\hat{I}_0(\hat{N},\hat{\theta}_x)$ for $\hat{N}=1$ as a function of $\hat{\theta}_x$. Right: The behavior of $\hat{W}_0$ as a function of $\hat{N}$. The blue solid line refers to direct computation of  Eq. (\ref{unduradintenorm3}) and Eq. (\ref{unduradWnorm3}). The orange circles refer to Eq. (\ref{unduradintenormlim1a}) and   Eq. (\ref{unduradWnormlim1a}). }
	\label{figure3}
\end{figure}

This expression is valid for any value of $\hat{N}$. An example is shown in Fig. \ref{figure3}.

We now turn to consider special asymptotes when $\hat{\alpha}_x \ne 0$ and $\hat{N}$ is small or large.

\subsection{Case for $\hat{N} \ll 1$}

\begin{figure}
	\centering
	\includegraphics[width=0.5\textwidth]{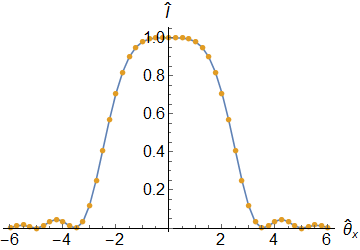}% \hfil%
    \includegraphics[width=0.5\textwidth]{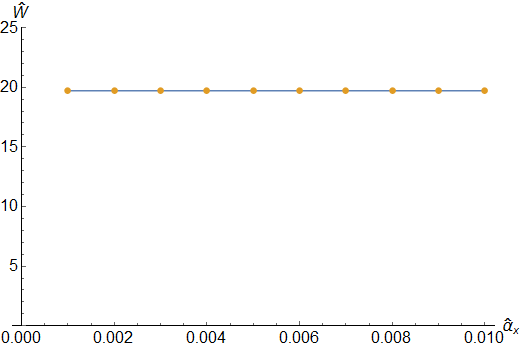}% \hfil%
	\caption{Left: $\hat{I}$ for $\hat{N} \ll 1$ and $\hat{\alpha}_x \ll 1$ as a function of $\hat{\theta}_x$. Right: The behavior of $\hat{W}_0$ as a function of $\hat{\alpha}_x$ for $\hat{N} = 10^{-4}$ . The blue solid line refers to direct computation of  Eq. (\ref{unduradintenorm3}) and Eq. (\ref{unduradWnorm3}). The orange circles refer to Eq. (\ref{unduradintenormlim1b}) and   Eq. (\ref{unduradWnormlim1b}). }
	\label{figure33}
\end{figure}

If $\hat{\alpha}_x \ll 1$ one gets back the limiting case of Eq. (\ref{unduradintenormlim1a}) and Eq. (\ref{unduradWnormlim1a}) describing emission from a single particle moving along the $z$ axis:

\begin{eqnarray}
\hat{I} = \lim_{\hat{N} \rightarrow 0} \hat{I}_0(\hat{N},\hat{\theta}_x)  &&= \mathrm{sinc}^2\left[\frac{ \hat{\theta}_x^2}{4}\right] ~.
\label{unduradintenormlim1b}
\end{eqnarray}
\begin{eqnarray}
\hat{W} &&= \lim_{\hat{N} \rightarrow 0}  \hat{W}_0(\hat{N}) = 2\pi^2
\label{unduradWnormlim1b}
\end{eqnarray}
The behavior of $\hat{I}$ and $\hat{W}$ is illustrated in Fig. \ref{figure33}.

If $\hat{\alpha}_x \sim 1$ one has

\begin{figure}
	\centering
	\includegraphics[width=0.5\textwidth]{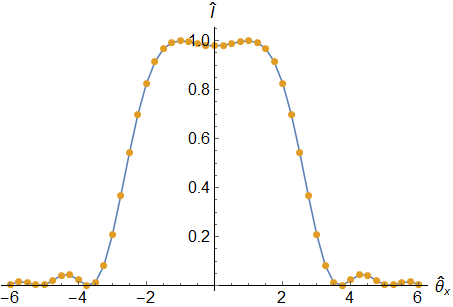}% \hfil%
    \includegraphics[width=0.5\textwidth]{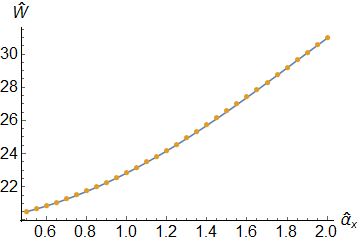}% \hfil%
	\caption{Left: $\hat{I}$ for $\hat{N} \ll 1$ and $\hat{\alpha}_x = 1$ as a function of $\hat{\theta}_x$. Right: The behavior of $\hat{W}_0$ as a function of $\hat{\alpha}_x \sim 1$ for $\hat{N} = 10^{-4}$ . The blue solid line refers to direct computation of  Eq. (\ref{unduradintenorm3}) and Eq. (\ref{unduradWnorm3}). The orange circles refer to Eq. (\ref{unduradintenorm3lim1}) and   Eq. (\ref{unduradWnorm2lim1}). }
	\label{figure34}
\end{figure}

\begin{eqnarray}
\hat{I}=   \hat{I}_1 && = \mathrm{sinc}^2\left[-\frac{\hat{\alpha}_x^2}{4}+\frac{ \hat{\theta}^2}{4}\right] ~
\label{unduradintenorm3lim1}
\end{eqnarray}
and

\begin{eqnarray}
\hat{W} = \hat{W}_1 &&  = 2 \pi \int_0^\infty d\hat{\xi} ~\hat{\xi} \mathrm{sinc}^2\left[-\frac{\hat{\alpha}_x^2}{4}+\frac{ \hat{\xi}^2}{4}\right] = \cr && 2 \pi^2 \left[1-\frac{4}{\pi \hat{\alpha}_x^2} + \frac{4}{\pi \hat{\alpha}_x^2} \cos\left(\frac{\hat{\alpha}_x^2}{2} \right) + \frac{1}{\pi} \mathrm{Si}\left(\frac{\hat{\alpha}_x^2}{2} \right)\right]~.
\label{unduradWnorm2lim1}
\end{eqnarray}
with $\mathrm{Si}$ the Sin Integral function. This limit corresponds to a red-shifted single-particle case, and is valid for increasing values of $\hat{\alpha}_x$, until $\hat{N} \hat{\alpha}_x^2 \ll 1$. The behavior of $\hat{I}$ and $\hat{W}$ is illustrated in Fig. \ref{figure34}. When $\alpha_x$ increases such that $\hat{N} \hat{\alpha}_x^2 \gtrsim 1$ or larger the general Eq. (\ref{unduradintenorm3}) and Eq. (\ref{unduradWnorm3}) must be used.

\subsection{Case for $\hat{N} \gg 1$}

\begin{figure}
	\centering
	\includegraphics[width=0.5\textwidth]{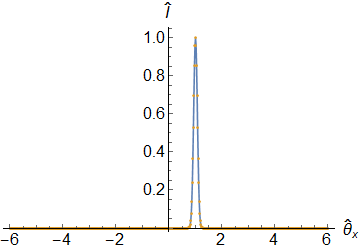}% \hfil%
    \includegraphics[width=0.5\textwidth]{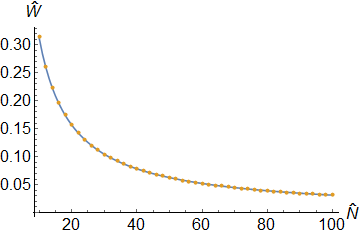}% \hfil%
	\caption{Left: $\hat{I}$ for $\hat{N} = 100$ and $\hat{\alpha}_x = 1$ as a function of $\hat{\theta}_x$. Right: The behavior of $\hat{W}_0$ as a function of $\hat{N}$ for $\hat{\alpha}_x = 1$ . The blue solid line refers to direct computation of  Eq. (\ref{unduradintenorm3}) and Eq. (\ref{unduradWnorm3}). The orange circles refer to Eq. (\ref{unduradintenorm3lim4}) and   Eq. (\ref{unduradWnorm3lim4}). }
	\label{figure34}
\end{figure}

In the opposite limiting case,  Eq. (\ref{unduradintenorm3}) and Eq. (\ref{unduradWnorm3}) always simplifies to

\begin{eqnarray}
\hat{I} &&=  \exp\left[-\hat{N}( {\hat{\theta}_x} - {\hat{\alpha}_x})^2\right]~.
\label{unduradintenorm3lim4}
\end{eqnarray}
\begin{eqnarray}
\hat{W} &&=  \frac{\pi}{\hat{N}}~.\cr &&
\label{unduradWnorm3lim4}
\end{eqnarray}

\section{Discussion and Conclusions}

In the previous Section we reviewed with a critical eye the theory in \cite{KITA}. As discussed, Eq. (\ref{xpowden2}) is nothing but Eq. (4) in reference \cite{KITA}, and the power is shown to be a function of the angle between the electron beam direction and the microbunching wavefront normal, that is $\vec{\eta}-\vec{\alpha}$ according to our notations. We thus conclude that the results presented in \cite{KITA} are correctly derived in the framework of the conventional theory of radiation from relativistic charged particles. After deriving Eq. (\ref{xpowden2}) we further analyzed the emission properties of a microbunched electron beam with a wavefront tilt.

Our main conclusion is that (see Eq. (\ref{unduradintenorm2})) the radiation emission from a microbunched electron beam with the microbunching wavefront tilted of an angle $\alpha_x$ with respect to the direction of motion is maximum is for $\hat{\theta}_x = \hat{\alpha}_x$ and red-shifted of $\hat{C} = \hat{\alpha}^2/2$ or, in dimensional units, $\theta_x = \alpha_x$ and $C = - \omega \alpha_x^2 /(2 c) = - k_w \alpha_x^2 \gamma^2/(1 +K^2/2)$. In other words, for $\alpha_x \ne 0$, the conventional theory predicts emission of radiation as in Fig. \ref{figure2}(top) compared to the case for $\alpha_x = 0$ in Fig. \ref{figure2}(bottom).

The experiment described in \cite{NUHN,NATAL} shows, at variance, a maximum radiation emission in the direction of motion, that is at $\theta_x=0$ and, still, red-shifted. It should be underlined that the experiment was performed with a helical undulator while our calculations were performed for a planar undulator. However, as discussed above, the dimensionless result in Eq. (\ref{unduradintenorm2}) remains valid also for the helical case. To our understanding, the only way of obtaining maximum radiation emission in the direction of motion would be a readjustment of the microbunching wavefront. However, according to conventional particle tracking, the direction of the microbunching wavefront is not influenced by the kick. Concerning this last point we would like to draw the reader's attention to reference \cite{L}, which deals with the issue of separation of circular and linear polarization components from a setup similar to that built at the LCLS (linear undulator followed by a kick and a helical radiator, without inverse tapering). The authors of \cite{L}, knew that coherent radiation emission is exponentially suppressed, unless the microbunching is directed along the velocity.  Therefore, they proposed a design  of an  isochronous  bending system based on the use of conventional particle tracking and  XFEL codes. For the European XFEL it requires about 87 m long of total length and consists of 33 magnets, including 8 dipoles, 9 quadrupoles, and 16 sextupoles. Such a system would allow for a rotation of the microbunching wavefront of an angle equal to the bending angle, thus yielding strong coherent emission at resonance in the direction of motion. In reality, no isochronous bending system was actually needed at the LCLS facility to achieve intense emission of coherent, highly circularly polarized radiation. However, at the LCLS, the radiation observed was red-shifted.

The LCLS crew tend to ascribe their observations \cite{NUHN,NATAL} to  microbunching wavefront readjusting due to the presence of FEL gain in the final radiator. This explanation is not convincing for us: in our view, the FEL gain in the final radiator should only be accounted for as a spurious effect. At the entrance of the radiator, the initial microbunching -calculated according to the usual particle tracking techniques-, is directed at an angle with respect to the velocity.  At this initial position, the microbunching is considerable even if not saturated. The readjusting of the microbunching direction  could only happen after smearing of the initial microbunching and development of new microbunching along the velocity direction, but this would require many gain lengths, and is impossile in a short radiator. It is also of fundamental importance to note that a simple "rotation" of the microbunching direction would not explain the observation of red-shifted radiation,  in contrast to the observations in \cite{NUHN,NATAL}.

We therefore conclude that the observations in \cite{NUHN,NATAL} are currently at odds with the present theoretical understanding of radiation emission from relativistic charged particles and a fundamental explanation of this effect needs to be provided\footnote{In \cite{OURT}  we proposed an explanation which is in contrast with the mainstream understanding of the coupling between dynamics and electrodynamics, but is nevertheless capable of explaining the effects treated in this paper from a fundamental viewpoint.}.

\end{document}